\documentclass[
reprint,
 amsmath,
 amssymb,
 aps,
]{revtex4-2}

\usepackage{graphicx}
\usepackage{dcolumn}
\usepackage{bm}
\usepackage{hyperref}
\usepackage{xcolor}
\usepackage{comment}
\mathchardef\mhyphen="2D

\begin{document}

\preprint{APS/123-QED}

\title{Angular momentum alignment-to-orientation conversion in the ground state of Rb atoms at room temperature}

\author{A. Mozers}
 \email{arturs.mozers@lu.lv}
\author{L. Busaite}
\author{D. Osite}
\author{M. Auzinsh}
\affiliation{
 Laser Centre, University of Latvia, Rainis Boulevard 19, LV-1586 Riga, Latvia
}

\date{\today}

\begin{abstract}
We investigated experimentally and theoretically angular momentum alignment-to-orientation conversion created by the joint interaction of laser radiation and an external magnetic field with atomic rubidium at room temperature. In particular we were interested in alignment-to-orientation conversion in atomic ground state. Experimentally the laser frequency was fixed to the hyperfine transitions of $D_1$ line of rubidium. We used a theoretical model for signal simulations that takes into account all neighboring hyperfine levels, the mixing of magnetic sublevels in an external magnetic field, the coherence properties of the exciting laser radiation, and the Doppler effect. The experiments were carried out by exciting the atoms with linearly polarized laser radiation. Two oppositely circularly polarized laser induced fluorescence (LIF) components were detected and afterwards their difference was taken. The combined LIF signals originating from the hyperfine magnetic sublevel transitions of $^{85}$Rb and $^{87}$Rb rubidium isotopes were included. The alignment-to-orientation conversion can be undoubtedly identified in the difference signals for various laser frequencies as well as change in signal shapes can be observed when the laser power density is increased. We studied the formation and the underlying physical processes of the observed signal of the LIF components and their difference by performing the analysis of the influence of incoherent and coherent effects. We performed simulations of theoretical signals that showed the influence of ground-state coherent effects on the LIF difference signal.
\end{abstract}

\maketitle

\section{\label{sec:level1}Introduction}

Laser radiation by its very nature has specially anisotropic electric field distribution. First, the laser beam defines a direction in space to which the vector of the electric field of the light is always perpendicular. So it means that light electric field is always in a plane perpendicular to the beam direction. In addition, laser radiation very often is polarized. For example, if it is polarized linearly, then there exists a direction perpendicular to the beam direction which defines direction of laser radiation polarization.

When such a radiation is absorbed by an ensemble of atoms, it creates the spatial anisotropy of angular momentum distribution in atoms. This angular momentum spatial distribution anisotropy has the same spatial symmetry as the electric field vector of the exciting light. 

When the laser beam is linearly polarized, it creates alignment of the angular momentum of atoms in the excited state. If the absorption is nonlinear, alignment is created in the ground state of the atoms as well. Angular momentum alignment can be imagined as a double-headed arrow. If the angular momentum of the atoms is aligned along the quantization axis it is called the longitudinal alignment. In the case of longitudinal alignment the populations of magnetic sublevels with quantum number $+m_F$ and $-m_F$ are equal, but the population is different for different \(|{m_F}|\) states. But if the angular momentum is aligned perpendicularly to the quantization axis (it is called transverse alignment), it means that there is a coherence created between magnetic sublevels with quantum numbers that differ by $\Delta{m_F}= \pm 2$, for details see~\cite{Auzinsh:2010book, Auzinsh:2005MolPol}.

In a similar way we can introduce longitudinal and transverse orientation of angular momentum. Usually orientation can be created by a circularly polarized laser excitation. In the case of orientation of the angular momentum, the spatial distribution can be represented symbolically by a single-headed arrow, and in the case of longitudinal orientation the magnetic sublevels with quantum numbers $+m_F$ and $-m_F$ in general have different populations. However, the case of transverse orientation corresponds to coherence between magnetic sublevels with values that differ by $\Delta{m_F}= \pm 1$ \cite{Auzinsh:2010book, Auzinsh:2005MolPol}.

Changes in the fluorescence polarization, for example, depolarization of laser induced fluorescence in the magnetic field -- Hanle effect \cite{Moruzzi1991} or the rotation of the plane of polarization of laser radiation that propagates in the gas of atoms~\cite{Budker:2002}, have broad range of applications, for example, these effects can be used to measure the magnetic field. Other manifestations of magneto-optical resonances include electromagnetically induced transparency~\cite{Harris:1997}, information storage in atoms~\cite{Phillips:2001, Liu:2001}, atomic clocks~\cite{Knappe:2005}, range of optical switches~\cite{Yeh:1982} and filters~\cite{Cere:2009}, as well as optical isolators can be designed~\cite{Weller:2012} based on these effects.

At some specific conditions the alignment created by a linearly polarized laser excitation can be converted to orientation. For that some additional perturbation is needed. For example, it can be magnetic field gradient~\cite{Fano:1964} or anisotropic collisions~\cite{Lombardi:1967,Rebane:1968,Manabe:1981}.

This process is called alignment-to-orientation conversion (AOC)~\cite{Auzinsh:2005MolPol}. 

Interaction with an electric field also can produce orientation from an initially aligned 
population~\cite{Lombardi:1969}.

A magnetic field alone cannot create atomic angular momentum orientation from an initially aligned ensemble. because it is an axial field that has even parity or it is symmetric under reflection in the plane perpendicular to the field direction. 

However, if the strength of magnetic field interaction with an atom is comparable with hyperfine interaction in the atom the joint action of both interactions can cause alignment to orientation conversion~\cite{Alnis:2001}.

\begin{figure}[h!]
	\includegraphics[width=\linewidth]{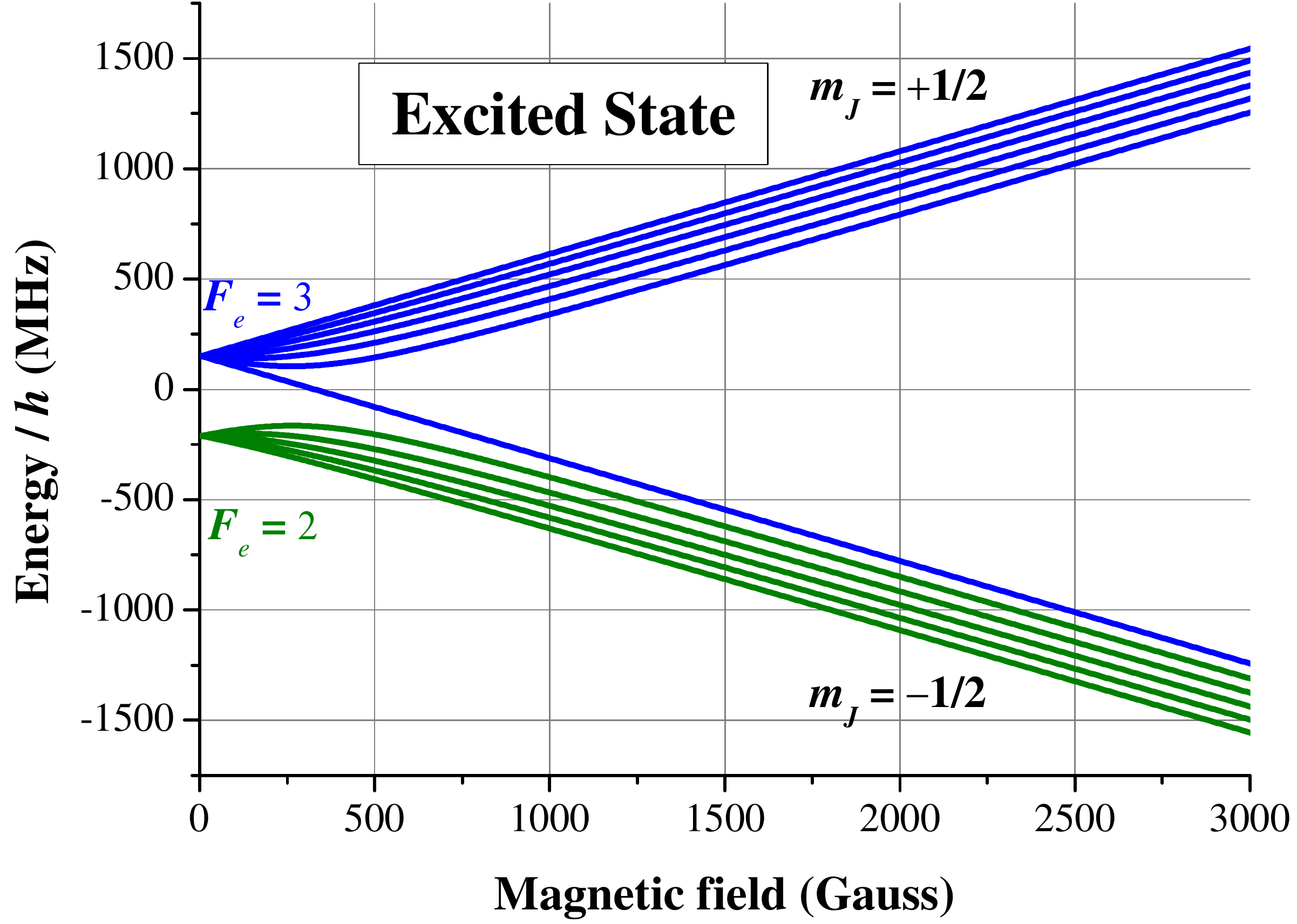}
    \caption{\label{fig:im1}Frequency shifts of the magnetic sublevels $m_F$ of the excited-state fine-structure level 5$^2$P$_{1/2}$ as a function of magnetic field for $^{85}$Rb. Zero frequency shift corresponds to the excited-state fine-structure level 5$^2$P$_{1/2}$.
    }
\end{figure}

\begin{figure}[h!]
	\includegraphics[width=\linewidth]{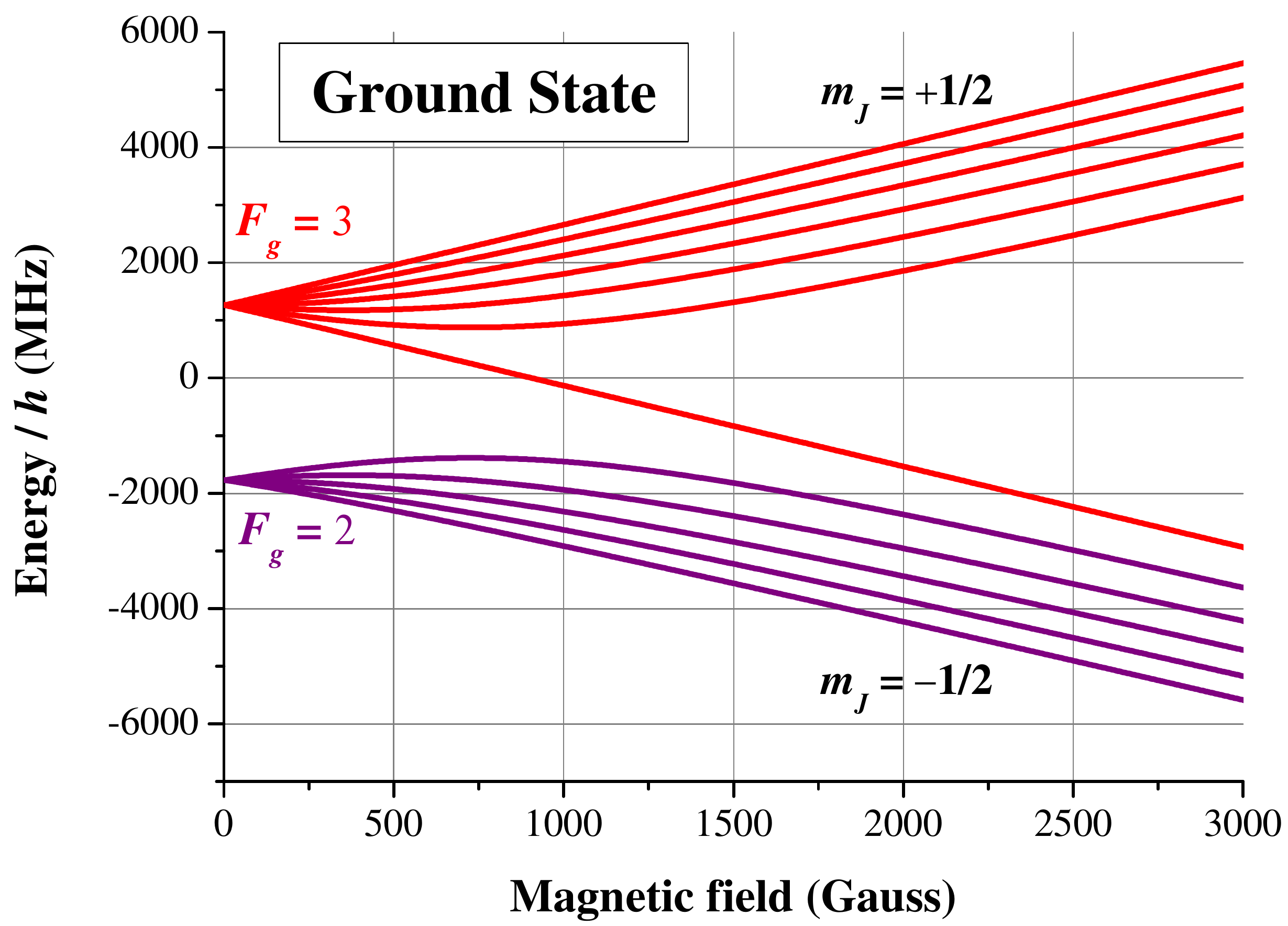}
    \caption{\label{fig:im2}Frequency shifts of the magnetic sublevels $m_F$ of the ground-state fine-structure level 5$^2$S$_{1/2}$ as a function of magnetic field for $^{85}$Rb. Zero frequency shift corresponds to the ground-state fine-structure level 5$^2$S$_{1/2}$.
    }
\end{figure}

At the intermediate magnetic field strength the hyperfine interaction can cause a nonlinear dependence of the energies of the magnetic sublevels on the magnitude of the magnetic field---the nonlinear Zeeman or hyperfine Paschen--Back effect (see Fig.~\ref{fig:im1} and Fig.~\ref{fig:im2}). If, in addition, we have such a linearly polarized excitation radiation that it simultaneously contains linear ($\pi^0$) and circular ($\sigma^{\pm}$) polarization components in the reference frame defined by the magnetic field direction  (see Fig.~\ref{fig:im3}), then $\Delta m_F=1$ coherences can be created, which leads to  the breaking of the angular momentum spatial distribution symmetry~\cite{Alnis:2001} and creates the angular momentum orientation in a direction transverse to the magnetic field direction.

AOC in an external magnetic field was first studied theoretically for cadmium~\cite{Lehmann:1964} and sodium~\cite{Baylis:1968}, and observed experimentally in cadmium~\cite{Lehmann:1969} and in the $D_2$ line of rubidium atoms~\cite{Krainska-Miszczak:1979}. Also the conversion in the opposite sense---conversion of an oriented state into an aligned---is possible~\cite{Weiss:1978}. 

It can be concluded that in general the action of external perturbations can break the symmetry of the angular momenta distribution and allow the linearly polarized exciting radiation to produce orientation, which is manifested by the presence of circularly polarized fluorescence.
 
At the beginning, AOC caused by the joint action of external magnetic field and internal hyperfine interaction was studied in the rubidium atoms in their excited state. An excitation with weak laser radiation in the linear absorption regime was used~\cite{Alnis:2001}. 

The magnetic sublevels of the excited-state angular momentum hyperfine levels in Rb atoms in an external magnetic field start to be affected by the nonlinear Zeeman effect already at moderate field strengths of several tens of Gauss. It should be noted that at this magnetic field strength the ground-state Zeeman effect is still close to linear. 

However, many practical and experimental applications require excitation with higher laser power density, in which case the absorption becomes nonlinear.

Detailed study of alignment-to-orientation conversion in an excited state of Rb atoms in the case of nonlinear absorption recently was carried out in \cite{Mozers:2015}. 

Strongest alignment-to-orientation conversion happens at the magnetic field strength at which coherently excited magnetic sublevel pairs undergo level crossing due to nonlinear magnetic sublevel splitting~\cite{Alnis:2001, Mozers:2015}.
Such level crossings do not happen in the ground state of alkali atoms. For this reason usually it is not analyzed if the AOC can still happen in the ground state. At the same time it is known, see for example \cite{Alnis:2001}, that although level crossing strongly enhances AOC signals, conversion can also happen without level crossing. The only requirement is that excitation conditions are such that between certain magnetic sublevels coherences can be created, for example due to finite absorption line-width and width of the laser radiation spectral profile.

The aim of this study is to examine AOC processes in detail and to deconvolute AOC processes caused by different effects -- processes in excited state, processes in the ground state, changes in the absorption probabilities caused by the magnetic sublevel scanning in the external magnetic field and the effect that several isotopes of an atom can interact with the same laser radiation simultaneously.

Although AOC effects usually are small, especially those that are caused by the ground state of atoms it is important to have a clear understanding of them not only due to theoretical interest, but also due to practical and fundamental applications where they can play an important role. 

Not to go into great detail, we will mention here just one, but fundamentally interesting, example.

Experiments with atoms and molecules allow to conduct a search of the permanent electric dipole moment (EDM) of the electron with high sensitivity~\cite{khriplovich1997cp}.

Various physical effect can contaminate EDM search signals and lead to systematic errors. One such source for signal contamination is alignment-to-orientation conversion, see \cite{Auzinsh:2006} and references cited therein. For that reason ground-state level participation in AOC processes and measured signals are important to take into account.

All signals obtained from experiments were analysed by a numerical theoretical model based on optical Bloch equations (see section ``Theoretical model"). We carried out experiments where two oppositely circularly polarized LIF components and afterwards their difference were observed (section ``Experiment"). Experimental results clearly show AOC happening for various laser frequencies as well as change in signal shapes is observed when the measurement of laser power dependence is performed. A detailed explanation of the observed experimental signals is provided in section ``Results \& Discussion".

\section{Theoretical Model}

Prior to the experimental  measurement, we made an assessment which hyperfine transitions in Rb atoms are the most suitable to detect the alignment-to-orientation conversion in the atomic ground state. To estimate the expected signal strength and to analyze the experiemental signals, we use the Liouville or optical Bloch equations (OBEs) for the density matrix $\rho$.
The atomic density matrix will be written in a basis defined by the whole manifold of hyperfine levels in the ground and excited state: $\vert \xi, F_i, m_{F_i} \rangle$, where $F_i$ refers to the quantum number of the hyperfine angular momentum in the ground ($i = g$) or the excited ($i = e$) state, $m_{F_i}$ denotes the respective magnetic quantum number and $\xi$ represent all the other quantum numbers which are not essential for the current study.

The time evolution of the density matrix is described by the optical Bloch equations~\cite{Stenholm:2005}
\begin{equation}\label{eq:liouville}
	i\hbar\frac{\partial \rho}{\partial t} = \left[\hat H,\rho\right] + i\hbar\hat R\rho,
\end{equation}
where $\hat H$ is the total Hamilton operator of the system and $\hat R$ is the relaxation operator. The full Hamiltonian can be written as $\hat H = \hat H_0 + \hat H_B + \hat V$, where $\hat H_0$ is unperturbed system Hamiltonian, $\hat H_B$ describes the interaction with the external magnetic field, and $\hat V = -\hat{\mathbf{d}}\cdot \mathbf{E}(t)$ is the operator which describes atom -- laser field interaction in the electric dipole approximation. The operator  includes electric field of excitation light $\mathbf{E}(t)$ and an electric dipole operator $\hat{\mathbf{d}}$.

The general OBEs \eqref{eq:liouville} can be transformed into explicit rate equations for the Zeeman coherences within the ground ($\rho_{g_ig_j}$) and excited ($\rho_{e_ie_j}$) states by applying the rotating-wave approximation, averaging over and decorrelating from the stochastic phases of laser radiation, and adiabatically eliminating the optical coherences \cite{Blushs:2004, Auzinsh:2009, Auzinsh:2013,Auzinsh:2015}:
\begin{subequations} \label{eq:zc}
\begin{align}
\frac{\partial \rho_{g_ig_j}}{\partial t} =& \left(\Xi_{g_ie_m} + \Xi_{g_je_k}^{\ast}\right)\sum_{e_k,e_m}d_{g_ie_k}^\ast d_{e_mg_j}\rho_{e_ke_m} - \nonumber\\ -& \sum_{e_k,g_m}\Big(\Xi_{g_je_k}^{\ast}d_{g_ie_k}^\ast d_{e_kg_m}\rho_{g_mg_j} + \nonumber\\ +& \Xi_{g_ie_k}d_{g_me_k}^\ast d_{e_kg_j}\rho_{g_ig_m}\Big) -  i \omega_{g_ig_j}\rho_{g_ig_j} -\nonumber\\+&  \sum_{e_ke_l}\Gamma_{g_ig_j}^{e_ke_l}\rho_{e_ke_l} - \gamma\rho_{g_ig_j}  + \lambda\delta(g_i,g_j) \label{eq:zcgg} \\
\frac{\partial \rho_{e_ie_j}}{\partial t} =& \left(\Xi_{g_me_i}^\ast + \Xi_{g_ke_j}\right)\sum_{g_k,g_m}d_{e_ig_k} d_{g_me_j}^\ast\rho_{g_kg_m} -\nonumber\\-& \sum_{g_k,e_m}\Big(\Xi_{g_ke_j}d_{e_ig_k} d_{g_ke_m}^\ast\rho_{e_me_j} + \nonumber\\ +& \Xi_{g_ke_i}^\ast d_{e_mg_k} d_{g_ke_j}^\ast\rho_{e_ie_m}\Big) -  i \omega_{e_ie_j}\rho_{e_ie_j} -\nonumber\\ - & (\Gamma + \gamma)\rho_{e_ie_j}. \label{eq:zcee}
\end{align}
\end{subequations}
The equation Eq. \eqref{eq:zcgg} describes time evolution of the populations and Zeeman coherences in the manifold of magnetic sublevels of all the hyperfine levels in the atomic ground state. The first term describes the repopulation of the ground state and the creation of Zeeman coherences due to induced transitions, $\Xi_{g_ie_m}$ and $ \Xi_{g_je_k}^{\ast}$ represents the atom-laser field interaction strength. The matrix element $d_{g_ie_j} = \langle {g_i} | {\bf\hat d} \cdot {\bf e} | {e_j} \rangle$ is the dipole transition matrix element for transition between hyperfine level magnetic sublevels $g_i$ and $e_j$ when excited by laser radiation with polarization $\bf e$. The second term denotes the changes in ground-state Zeeman sublevel population and the creation of ground-state Zeeman coherences due to light absorption. The third term describes the destruction of the ground-state Zeeman coherences by the external magnetic field and hyperfine splitting, where $\omega_{g_ig_j}$ is the energy difference between magnetic sublevels $\vert i\rangle$ and $\vert j\rangle$. The fourth term describes the repopulation and transfer of excited-state coherences to the ground state due to spontaneous transitions.
The transtions are closed within hyperfine structure, so that $\sum_{e_ie_j}\Gamma^{e_ie_j}_{g_ig_j} = \Gamma$.
The fifth and sixth terms in (\ref{eq:zcgg}) describe the transit relaxation in the ground state. The fifth term accounts for the process of thermal motion in which atoms leave the region of atom interaction with a spatially restricted laser beam. The rate of this process is characterized by the constant $\gamma$. It is assumed that the atomic equilibrium density matrix outside the interaction region is the unit matrix $\hat 1$ divided by the total number of the magnetic sublevels $n_g$ in the ground-state hyperfine manifold. Therefore repopulation rate is connected to transit relaxation rate $\gamma$ as $\lambda = \gamma / n_g$, where $n_g$ is the total number of the magnetic sublevels in the ground state. The transit relaxation rate $\gamma$ can be roughly estimated as the inverse to the average time that atoms spend in the laser beam when they are traversing it due to thermal motion.

Similarly to the Eq. \eqref{eq:zcgg}, in Eq. \eqref{eq:zcee} the first term denotes the changes in the excited state density matrix caused by the light absorption, the second term describes induced transitions to the ground state, the third stands for the destruction of the excited-state Zeeman coherences in the external magnetic field and due to hyperfine splitting, where $\omega_{e_ie_j}$
is the splitting of the excited-state Zeeman sublevels. Finally, the fourth term describes the combined rate of spontaneous decay and transit relaxation (atoms are leaving the region where they interact with laser radiation due to thermal motion) of the excited state.

The atom-laser radiation interaction strength $\Xi_{g_ie_j}$, used in \eqref{eq:zc}, is expressed as:
\begin{equation}\label{eq:xi}
\Xi_{g_ie_j} = \frac{\Omega_{R}^{2}}{\frac{\Gamma+\gamma+\Delta\omega}{2}+\dot\imath\left(\bar\omega-\mathbf{k}_{\bar\omega}\cdot\mathbf{v}
	+\omega_{g_ie_j}\right)},
\end{equation}
where $\Omega_R$ is the reduced Rabi frequency, $\Omega_R^2$ being proportional to the laser power density. $\Delta\omega$ is the finite spectral width of the exciting radiation, $\bar \omega$ is the central frequency of the exciting radiation, $\mathbf{k}_{\bar\omega}$ the wave vector of exciting radiation, and $\mathbf{k}_{\bar\omega}\mathbf{v}$ is the Doppler shift experienced by an atom moving with velocity $\mathbf{v}$.

As far as the experiments were conducted at continuous wave excitation conditions, we are interested in the stationary solution of the equations \eqref{eq:zc} 
\begin{equation}\label{eq:steady}
	\frac{\partial \rho_{g_ig_j}}{\partial t} = \frac{\partial \rho_{e_ie_j}}{\partial t} = 0,
\end{equation}
reducing the differential equations \eqref{eq:zc} to the system of linear equations. The solution of the system yields density matrices for the ground and excited states.

The observed fluorescence intensity of polarization $\mathbf{e}_{fl}$ can be then calculated from excited state density matrix elements as
\begin{equation}\label{eq:fluorescence}
	I_{fl}(\mathbf{e}_{fl}) = \tilde{I}_0\sum\limits_{g_i,e_j,e_k} d_{g_ie_j}^{\ast(ob)}d_{e_kg_i}^{(ob)}\rho_{e_je_k},
\end{equation}
where $d_{e_ig_j}^{(ob)}$ are the dipole transition matrix elements for the radiation with specific polarization observed in a chosen direction. $\tilde{I}_0$ is the constant of proportionality.

The thermal motion of the atoms was accounted for by signal averaging over the thermal velocity distribution of atoms. This averaging was performed  by solving the Eqs. \eqref{eq:zc} for each velocity group, accounting for relative probability for atoms to have this velocity and averaging the fluorescence intensity \eqref{eq:fluorescence} over this distribution.

To simulate expected signals and to fit experimental results, as the first approximation we estimated the values of several parameters.

The transit relaxation rate can be estimated from the mean thermal velocity $v_{th}$ of the atoms projected onto the plane perpendicular to the laser beam and from the laser-beam diameter $d$:
\begin{equation}\label{eq:gamma}
	\gamma = \frac{v_{th}}{d},
\end{equation}

For $d = 1400~\mu$m and $T = 293$~K we estimate $\gamma = 2 \pi \cdot (0.019$~MHz$)$.

The reduced Rabi frequency is estimated as
\begin{equation}\label{eq:Rabi}
	\Omega_R = k_{R}\frac{\vert\vert d\vert\vert\cdot\vert\varepsilon\vert}{\hbar} = k_{R}\frac{\vert\vert d\vert\vert}{\hbar}\sqrt{\frac{2I}{\epsilon_0 n c}},
\end{equation}
where $k_R$ is a dimensionless fitting parameter,  $\vert\vert d\vert\vert~=~4.231ea_{0}$ \cite{Auzinsh:2010book} is the reduced dipole matrix element for $D_1$ transition, where $e$ is the electron charge and $a_0$ is the Bohr radius~\cite{Auzinsh:2009b}, 
$I$ is the power density (directly related to the amplitude of the electric field $\vert\varepsilon\vert$), $\epsilon_0$ is the electric constant, $n$ is the refractive index, and $c$ is the speed of light. 

In practise, the power density $I$ is not constant across the laser beam, so that the estimation of the parameter $k_R$ is not straightforward. The theoretical model uses a constant value for power density instead of actual power distribution. 
As the power density is increased, $\Omega_R$ cannot be related to the square root of the power density $I$ by the same constant $k_R$ as for the lower power densities \cite{Fescenko:2012, Auzinsh:2015}, if one merely assumes that the laser power density distribution within the beam is Gaussian.

This leads to the more complex relationship between $I$ and $\Omega_R$ which has a simple explanation.
Our experiment was performed in the regime of nonlinear absorption, which leads to strong depletion of ground-state population for large laser power densities.
For low laser power density, the ground-state population is only slightly changed even at the center of the beam, where the light is most intense.
However, when the laser power is increased, the atoms in the center of the beam are more actively excited, leaving a low ground-state population in the center of the beam.
When the laser power density is increased even more, the region of population depletion expands to the ``wings'' of the Gaussian power density distribution.

Due to this spatially dependent population depletion, the main contribution to the signal for weaker laser radiation comes from the central parts of the laser beam where the power density is the highest, and the theoretical proportionality of $\Omega_R$ to the square root of power density continues to hold. However, for stronger laser radiation power density, the peripheral parts of the laser beam, where power density is lower, start to play a larger role in the absorption process, because ground-state population there is more significant than at the center of the beam.
Therefore, when increasing the laser power density, the different parts of the laser beam play a dominant part in the absorption process, and it should be related to the Rabi frequency $\Omega_R$ in the theoretical model. We account for this effect by adjusting the value of the coefficient $k_R$ in the theoretical model to achieve better correspondence between the experimental measurements and theoretical calculations.

An appropriate estimate of the spectral width used in the theoretical model was found to be $\Delta\omega = 2\pi\cdot(2$~MHz$)$, which is close to the value given by the manufacturer of the laser.

\section{Experiment}

\begin{figure}[ht!]
    \centering
	\includegraphics[width=\linewidth]{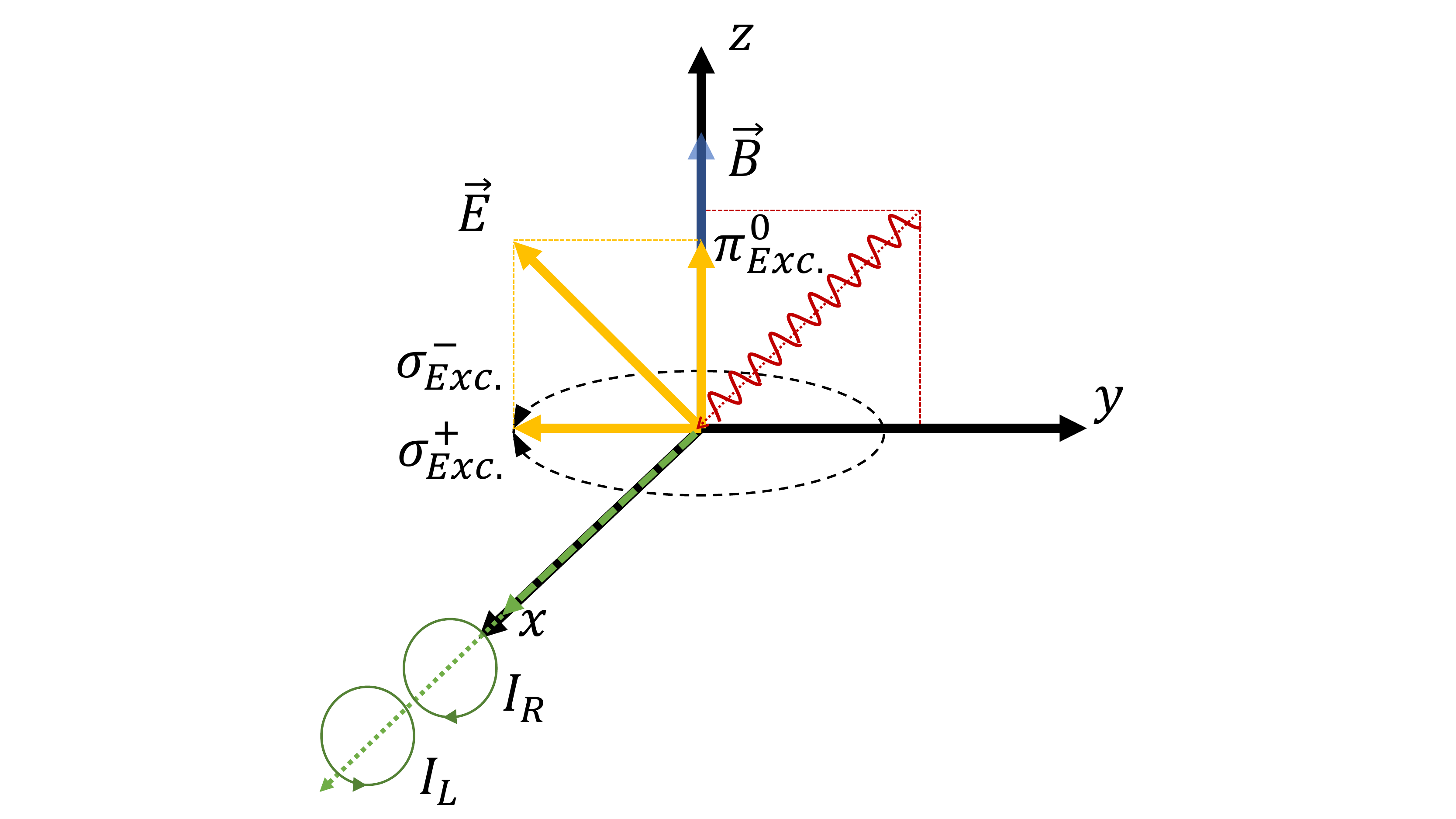}
	\centering
    \caption{\label{fig:im3}Excitation and observation geometry. The linearly polarized excitation laser light $\vec E$ can be split into two circularly polarized excitation light components $\sigma^\pm_{Exc.}$ and one linearly polarized excitation light component $\pi^0_{Exc.}$.}
\end{figure}

The experiments were performed on atomic rubidium vapor at room temperature. The cylindrical (diameter 25mm, length 25mm) Pyrex atomic vapor cell with optical quality windows from Toptica AG contained the natural abundance of rubidium isotopes. The atoms were excited with linearly polarized light with its polarization vector \textbf{E} in the \textit{y-z} plane and in a $\pi/4$ angle with respect to the quantization axis $z$ defined by the external magnetic field \textbf{B} as shown in Figure~\ref{fig:im3}. The two circularly polarized laser-induced fluorescence (LIF) components $I_L$ and $I_R$ were observed along the $x$-direction. The LIF passed through a pair of convex lenses while the discrimination between $I_L$ and $I_R$ was achieved by changing the relative angle between the fast axis of a zero-order quarter-wave plate (Thorlabs WPQ10M-780) and the polarization axis of a linear polarizer (LPVIS050-MP). These optical elements were aligned in a lens tube while the rotation of the linear polarizer was achieved by a rotation mount (CLR1/M). 

An external cavity, grating-stabilized, tuneable, single-mode diode laser DL 100, produced by Toptica AG, with a wavelength of 794.98 nm (D$_1$ line of Rb) and a typical linewidth of a few MHz, was used in all the experiments. The DTC110 and DCC110 modules from Toptica AG were used for temperature and current control of the laser. During the experiments the laser frequency was fixed to a saturation absorption spectrum (SAS) signal coming from another atomic rubidium vapor cell, which was placed in a three-layer $\mu$-metal shield. The locking of the laser frequency was established using the SC110 and DigiLock modules and software by Toptica AG. By using this feedback controlled loop it was possible to lock the laser frequency to a particular peak of the SAS, i.e. to a particular hyperfine transition.

\begin{figure}
    \includegraphics[width=.9\linewidth]{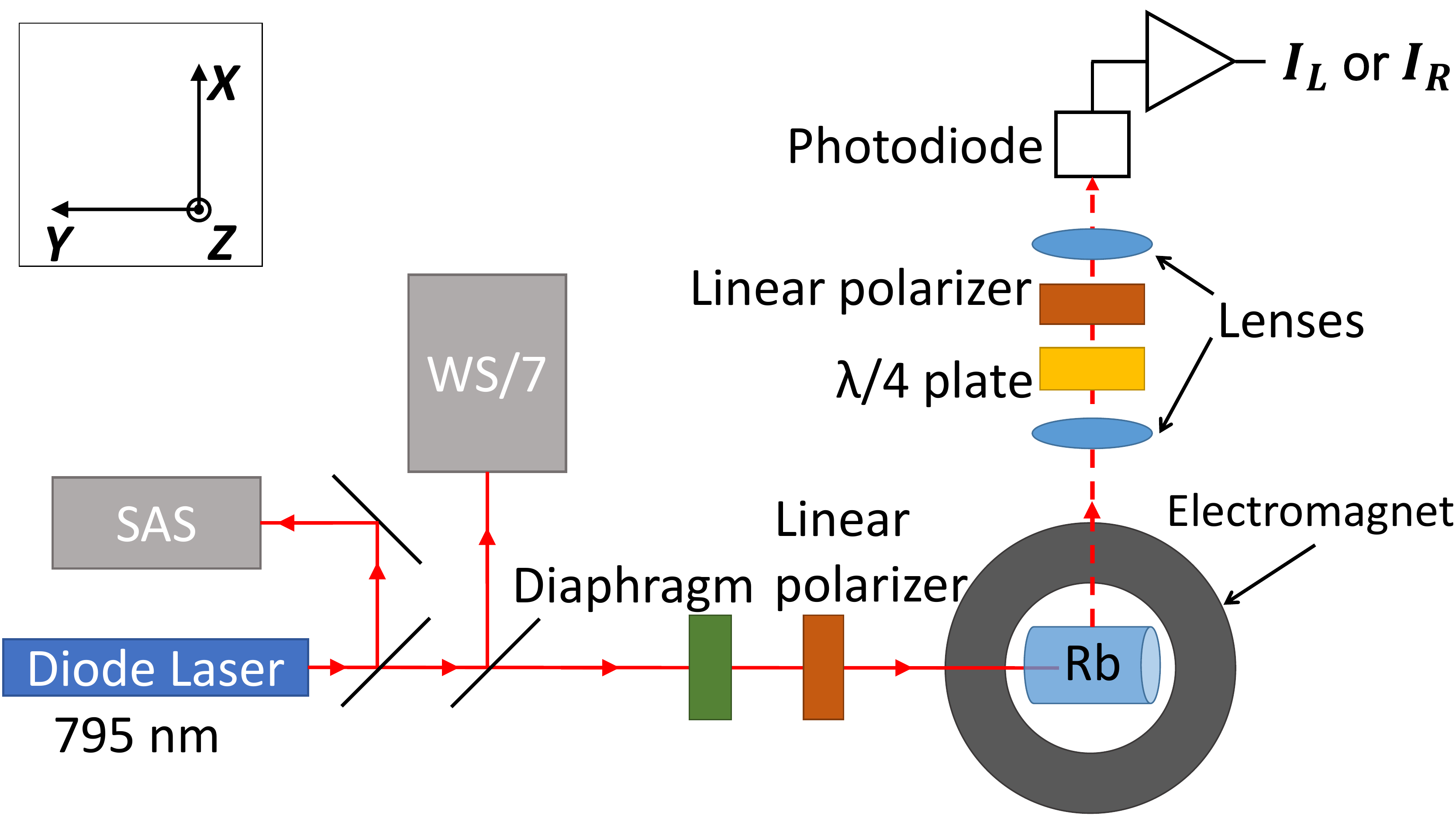}
    \centering
    \caption{\label{fig:im4} Side view of the experimental setup. The beam enters the coils at an angle of $45^\circ$ with respect to the $y$ axis in $yz$ plane (axes shown in inset).}
\end{figure}

Figure~\ref{fig:im4} shows a schematic of the experimental setup. The magnetic field was applied using an electromagnet with an iron core (diameter 10.0~cm), the separation between the surfaces of the poles was 4.3 cm. The inhomogeneity of the field in the center of the poles was estimated to be not more that 0.027\%. The current for the electromagnet was supplied by a KEPCO BOP20-10ML bipolar power supply and the symmetrical triangular current wave scan was generated by a function generator from TTi (TG 5011). The frequency of the magnetic field scan was 2.00 mHz with a maximum scan amplitude resulting in a magnetic field range from $-3100$ to $+3100$~G. 

The ellipticity of the laser beam was precluded by a diaphragm. The beam diameter was measured to be 1400 $\mu$m as full width at half maximum (FWHM) of the Gaussian fit by a beam profiler (Cohorent Inc. LASERCAM HR). The laser power was adjusted by a half-wave Fresnel rhomb retarder (FR600HM), followed by a linear polarizer (GTH10M). This enabled us to vary laser power values from 10 $\mu$W to 600 $\mu$W tranlating into laser power density from 0.36 mW/cm$^2$ to 28.6 mW/cm$^2$. The LIF was detected with a photodiode (Thorlabs SM1PD1A) which was placed at the end of and fixed into the observation lens tube. The LIF from each circularly polarized component was detected independently i.e. one at a time. The signal from the photodiode was amplified  by a transimpedance amplifier based on a TL072 op-amp (Roithner multiboard) with a gain of $10^6$, followed by a voltage amplifier with a gain of $10^4$. Every scan was acquired with the use of a digital oscilloscope Agilent DSO5014A and transferred to a PC with a minimum of 16 scans in total for each component.

Then the experimental signals of each LIF circularly component were averaged over multiple scans. To eliminate any residual asymmetry in the signal, an averaging over the negative and positive values of the magnetic field was performed. When comparing the experimental signals to theory, the constant background was subtracted, before the signals were normalized to the maximum of each component. The background was measured by blocking the laser beam and recording the signal from the photodiode.
In data processing we allowed the background value to vary for different laser power densities in order to achieve a better agreement between the experiment and the theory, but the variation of the background value never exceeded 3\%, which is within the measurement error of the measured background value.
As the LIF component signals were relatively large in comparison to their difference and circularity signals, a Savitzky–Golay smoothing filter~\cite{Savitzky-Golay} was applied to the LIF difference signals.

\section{Results \& Discussion}

\begin{figure}[ht!]
    \includegraphics[width=.9\linewidth]{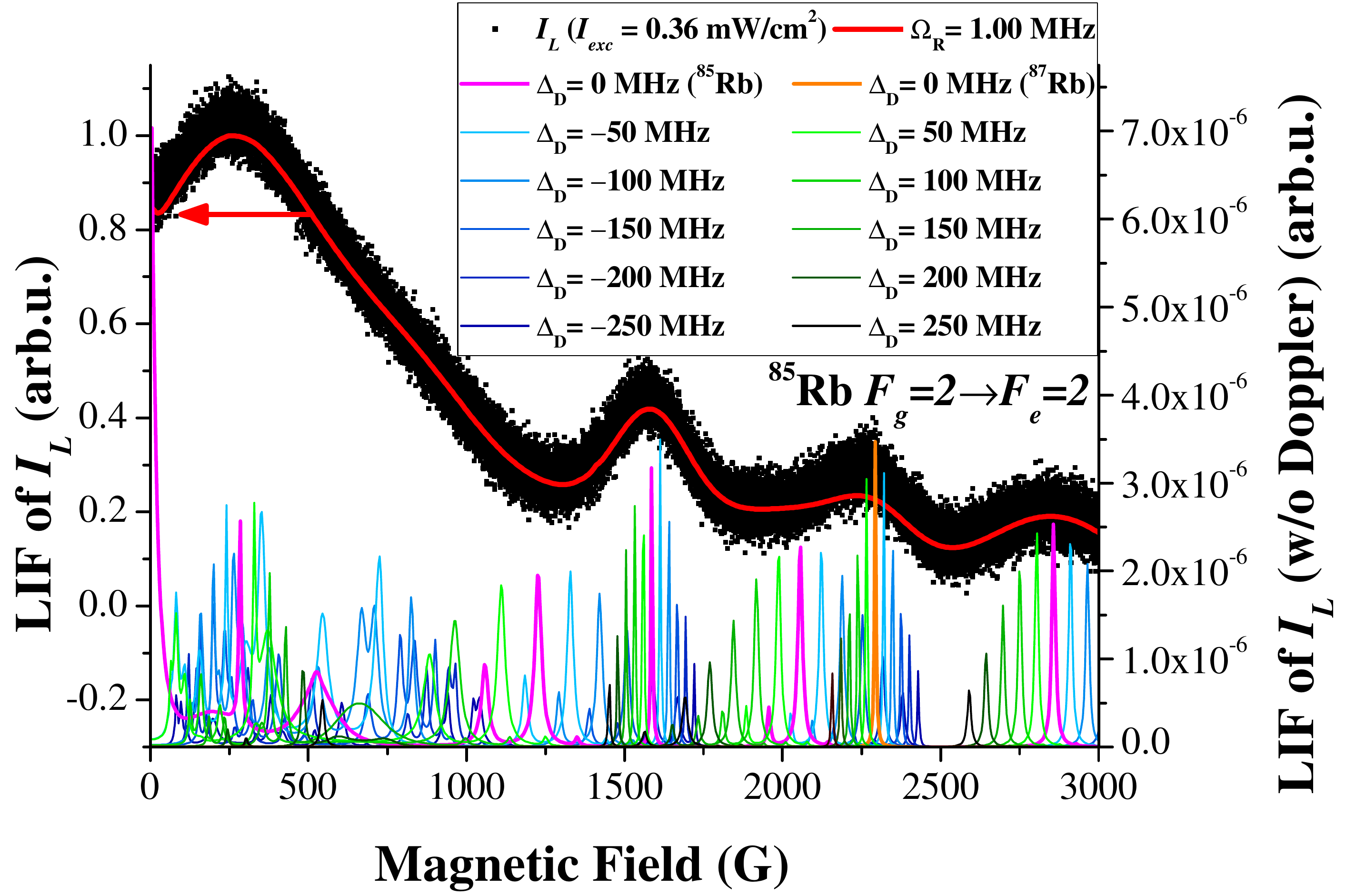}
    \centering
    \caption{\label{fig:im5}LIF of a single  circularly polarized light component ($I_L$): black dots -- experimental data; red line -- theoretical data (left axis); colored lines -- theoretical data without averaging over the Doppler profile (right axis). Magenta: central velocity group of $^{85}$Rb; orange: central velocity group of $^{87}$Rb; blue curves: negative velocity shift; green curves: positive velocity shift.}
\end{figure}

\begin{figure}[ht!]
    \includegraphics[width=.9\linewidth]{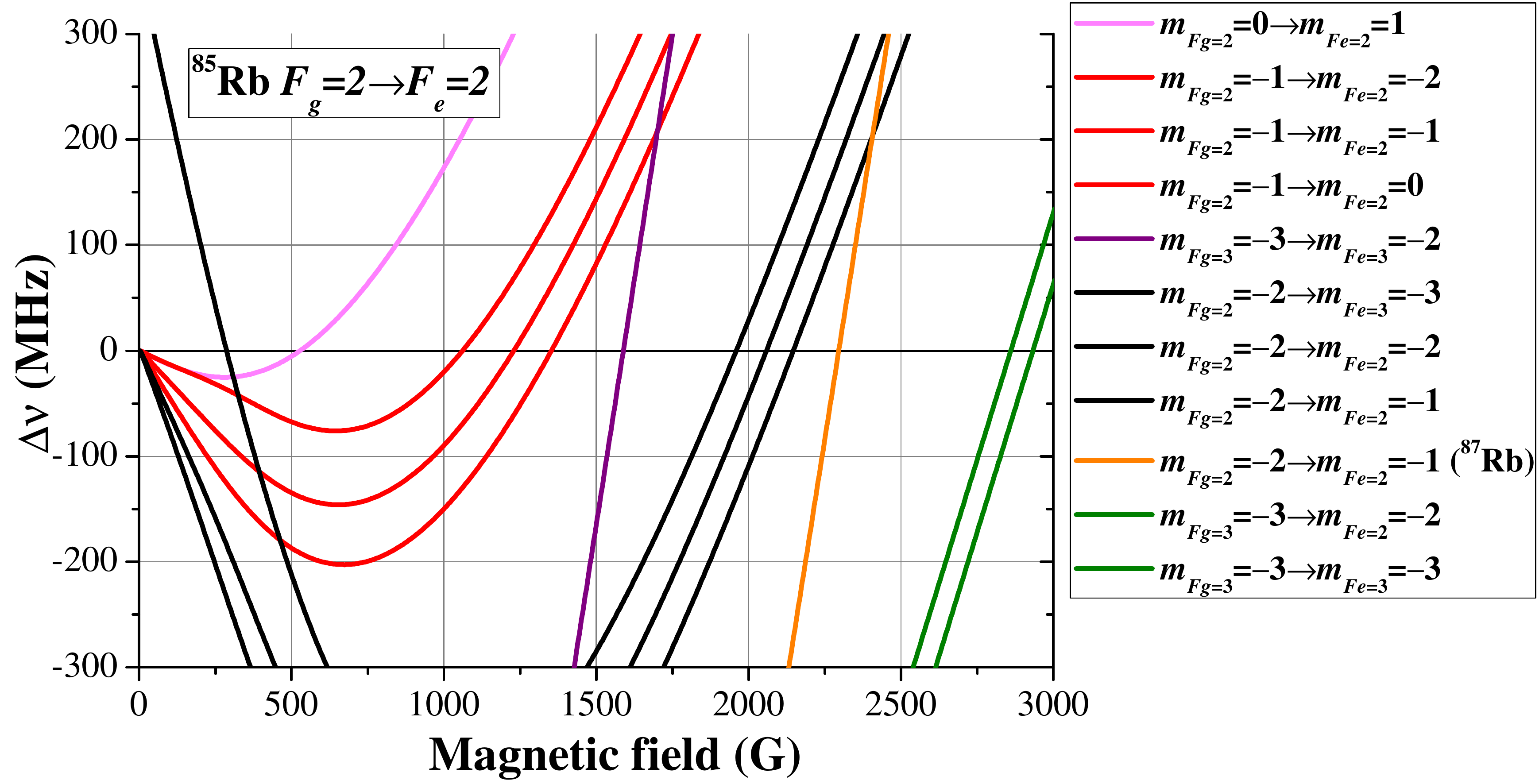}
    \centering
    \caption{\label{fig:im6}The colored lines represent the dependence of energy difference between various pairs of magnetic sublevels on magnetic field. $\Delta\nu=0$ corresponds to laser frequency equal to the $F_g=2\rightarrow F_e=2$ hyperfine transition of the $^{85}$Rb. (Only pairs crossing $\Delta\nu=0$ are shown) }
\end{figure}

Let us start with the analysis of the general structure of the observed signals. In this paper as an example we will show only one of the components ($I_L$), as the differences in the two oppositely circularly polarized LIF components are small and can be barely seen when the two observed circularly polarized LIF components are depicted side by side.
Figure~\ref{fig:im5} shows a typical result for the measurement of a single circularly polarized LIF component, when the laser frequency was locked to the $F_g=2\rightarrow F_e=2$ transition of the $^{85}$Rb. At zero magnetic field an initial relative minimum of the LIF signal can be observed.
The increase of the magnetic field lifts the degeneracy and the LIF signal rises because of other atoms with different velocity coming into resonance. The LIF signal starts to fall after approximately 250~G. The diminishing of the signal is caused by the same fact as the increase in signal, but now the contrary happens -- the nonlinear Zeeman effect both for the ground and excited state lead to a decrease in the number of atoms that interact with the laser light.
A pronounced feature can be seen at approximately 1500~G (and another at 2800~G) -- an increase in the LIF signal. This is caused by magnetic sublevels coming back into resonance with the excitation laser radiation. In particular, one can deduce exactly which magnetic sublevels are the ones that are interacting with the laser field from Figure~\ref{fig:im6}. Figure~\ref{fig:im6} shows the dependence of energy difference $\Delta\nu$ between pairs of magnetic sublevels on the external magnetic field. When $\Delta\nu$ is equal to 0, the energy difference between pairs of magnetic sublevels coincides with the laser frequency. Thus it can be easily deduced that the pair of magnetic sublevels corresponding to the increase in LIF at 1500~G are $m_{F_g=2}=-3\rightarrow m_{F_e=3}=-2$. 

Usually in theoretical simulations magnetic sublevels from only one isotope are being considered. For a complete understanding of the shapes of these signals one has to take into account both isotopes, as the magnetic field is large enough to bring magnetic sublevels into resonance originating from the hyperfine levels of $^{87}$Rb as well. In order to combine the LIF signals from both isotopes, the signals were weighed according to the difference in isotope abundance and line strength~\cite{Auzinsh:2010book}.

The width of these non-zero field structures in the observed signal can be attributed to the fact that the interacting atoms are in thermal motion, thus the Doppler effect plays a large role in the formation of these lineshapes. The curves below the experimental and isotopically combined LIF signal are the data from LIF signal simulations, where the averaging over the Doppler profile was omitted and a single velocity group was selected from the Doppler profile. Now the width of the shapes in the experimental data as well as in the simulated red curve in Figure ~\ref{fig:im5} can be interpreted as LIF coming from different velocity groups. The width of the narrow peaks appearing in the simulated LIF curves for single velocity groups is related to the combined width coming from the natural line-width and excitation laser line-width. The different relative amplitudes of these peaks in LIF signals from different velocity groups are related to transition probabilities between magnetic sublevels, i.e. when an external magnetic field is applied the wave functions of magnetic sublevels mix and their transition probabilities change~\cite{Sargsyan:14}. The summation over all of these LIF curves from different velocity groups (Doppler components) would yield the complete LIF simulated curve (Fig.~\ref{fig:im5} red).

A rather counter-intuitive feature can be noticed at approximately 1250~G. The zero velocity group LIF curve (Fig.~\ref{fig:im5} $\Delta_D=0$ MHz magenta) shows an increase in the signal and in Figure~\ref{fig:im6} magnetic sublevels $m_{F_g=2}=-1\rightarrow m_{F_e=2}=-1$ come into resonance with the laser. However, the averaging over the Doppler profile produces a minimum in the LIF signal. To understand the cause of such a peculiar feature in the observed signal, we show LIF from several Doppler components (velocity groups) in Figure~\ref{fig:im5} and energy difference $\Delta\nu$ between pairs of magnetic sublevels in Figure~\ref{fig:im6}. For a peak to appear in the observed signal the shift in transition frequency between two magnetic sublevels (change in $\Delta\nu$) should be larger than the change in the absolute value of the applied magnetic field (change in \textbf{B}). When a rather large change in \textbf{B} is necessary to achieve the same change in $\Delta\nu$, the Doppler components get spread out more and this flattens the overall signal, thus leading to a relative minimum in the observed signal.

\begin{figure}[ht!]
    \includegraphics[width=1\linewidth]{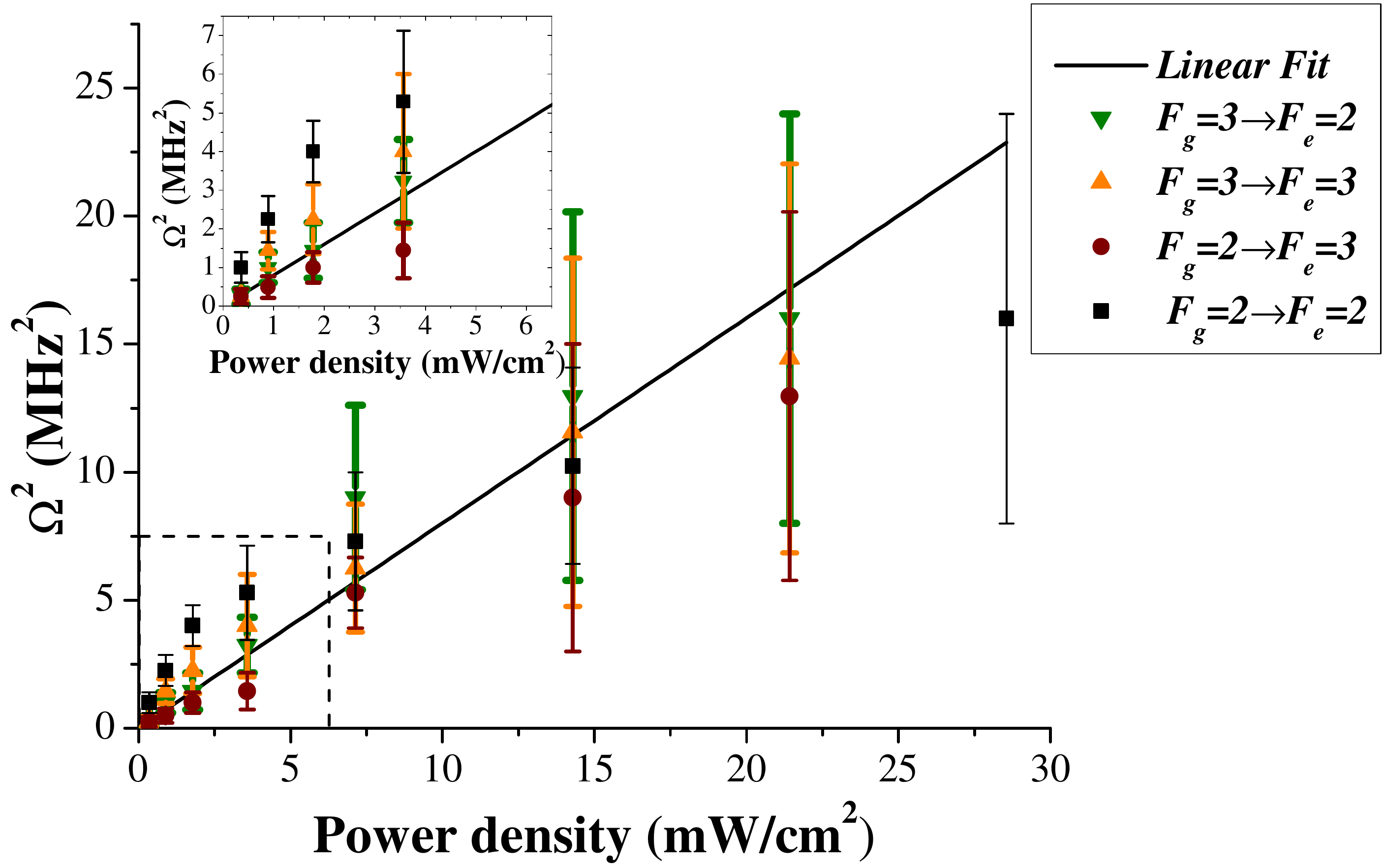}
    \centering
    \caption{\label{fig:im11}Dependence of the Rabi frequency squared $\Omega_R^2$ on the laser power density $I$ 
    together with a linear fit for all the fixed laser frequencies used in the experiment (colored data points).}
\end{figure}

We performed an analogous examination of all measured circularly polarized LIF signals for different exciting laser frequencies and power densities.
All experimentally obtained signals were fitted to simulated curves. Figure \ref{fig:im11} shows all of the Rabi frequency values (squared) obtained from the data fitting procedure vs laser power density. Different colors in Figure~\ref{fig:im11} correspond to different laser frequencies. The data points are in good agreement with a linear fit described by Eq.~\eqref{eq:Rabi}. We allowed the fitting parameter $k_R$ to vary in order to achieve better agreement between the experiment and the theory leading to a symmetric distribution of data points in Figure~\ref{fig:im11}.
It can be seen that the data points are in good agreement and within the margins of error.
Nevertheless, in closer examination the data points show a tendency to fall below the linear fit at laser power densities above 20 mW/cm$^2$. This happens because the theoretical model does not take into account the spatial distribution of the exciting optical field. The influence of different laser power density in different spatial positions on the fluorescence signal that causes the atoms to interact differently with the laser beam has been studied in~\cite{Kalnins2016}.

The LIF signal dependence on laser detuning was analysed in terms of the difference between the two observed LIF components defined as $I_L-I_R$. We show the difference signals as it depends only on the angular momentum transverse orientation besides the other measure of orientation -- circularity, which had much the same shape but which was slightly influenced by the dependence on the angular momentum alignment as well~\cite{Auzinsh:2005MolPol}.

\begin{figure*} 
    \includegraphics[width=1\linewidth]{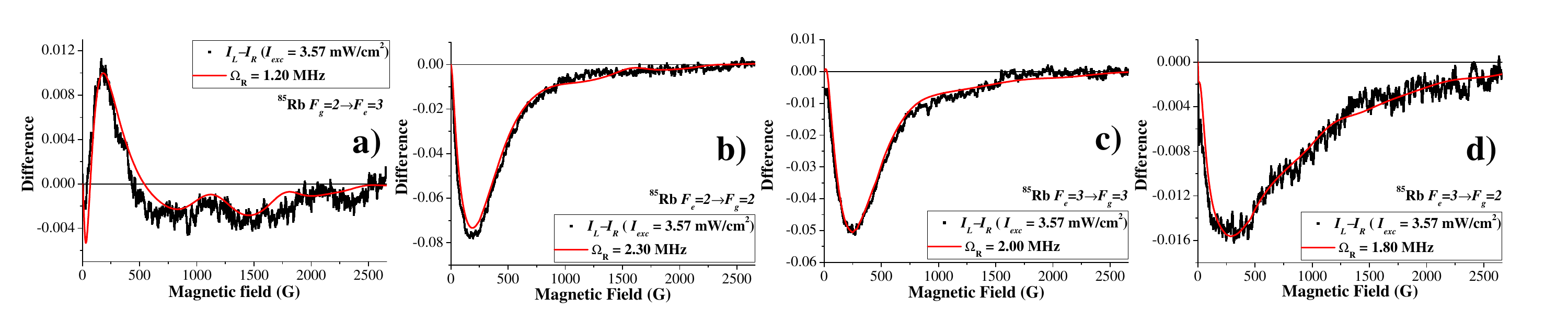}
    \centering
    \caption{\label{fig:im7}The dependence of the difference between the two LIF components on the hyperfine transitions that the laser frequency was fixed to. Laser frequency was fixed to: a) $F_g=2\rightarrow F_e=3$, b) $F_g=2\rightarrow F_e=2$, c) $F_g=3\rightarrow F_e=3$, d) $F_g=3\rightarrow F_e=2$ hyperfine transition of $^{85}$Rb. a)-d) laser frequencies are in descending order.}
\end{figure*}

Figure~\ref{fig:im7} shows the dependence of $I_L-I_R$ on the external magnetic field for different laser frequencies. As the laser frequency is increased (from a to d in Fig.~\ref{fig:im7}), a change in lineshapes can be observed. The circularity signal when the laser frequency was fixed to the $F_g=2\rightarrow F_e=3$ transition of the $^{85}$Rb changes from slightly positive ($\approx+0.5\%$) circularly polarized light to slightly negatively ($\approx-0.5\%$) polarized light. This indicates that the transverse orientation of the angular momentum of the ensemble of atoms also changes from slightly positive values to slightly negative values. In contrast, when the laser frequency was fixed to other hyperfine transitions, the sign of circularity did not change depending on the magnetic field. The largest circularity value of $\approx 4\%$ was observed when the laser frequency was set to the $F_g=2\rightarrow F_e=2$ transition of the $^{85}$Rb.

For the purpose of this study it is important to examine how the difference signal for the two circularly polarized fluorescence components (which in chosen geometry of excitation -- observation is directly proportional to the angular momentum transverse orientation) depends on the power density of the excitation radiation. This dependence serves as one of the indicators that helps to separate the effects of the atomic excited state that are present even at linear absorption region, from the ground-state effects that are intrinsically nonlinear with respect to the light intensity and do not manifest themselves at weak excitation laser power density.

\begin{figure*} 
    \includegraphics[width=.9\linewidth]{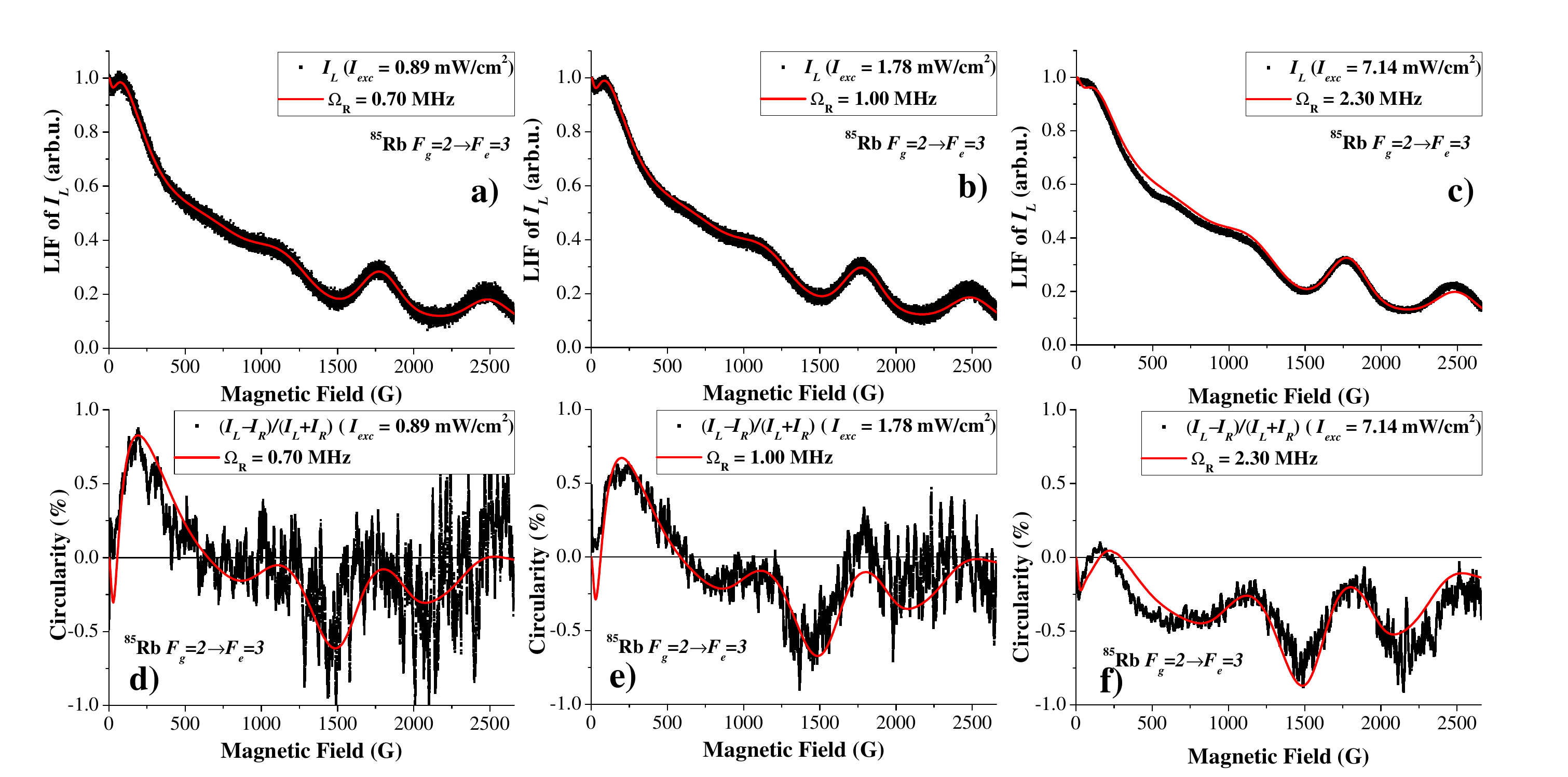}
    \centering
    \caption{\label{fig:im8}First row a)-c) shows the signal of a single circularly polarized LIF component $I_L$ dependence on laser power density. Second row d)-f) shows the corresponding circularity dependence on laser power density for the case when the laser frequency was fixed to the $F_g=2\rightarrow F_e=3$ hyperfine transition of $^{85}$Rb. Black dots: experimental data; red curve: theoretical calculation.}
\end{figure*}

\begin{figure} 
    \includegraphics[width=1\linewidth]{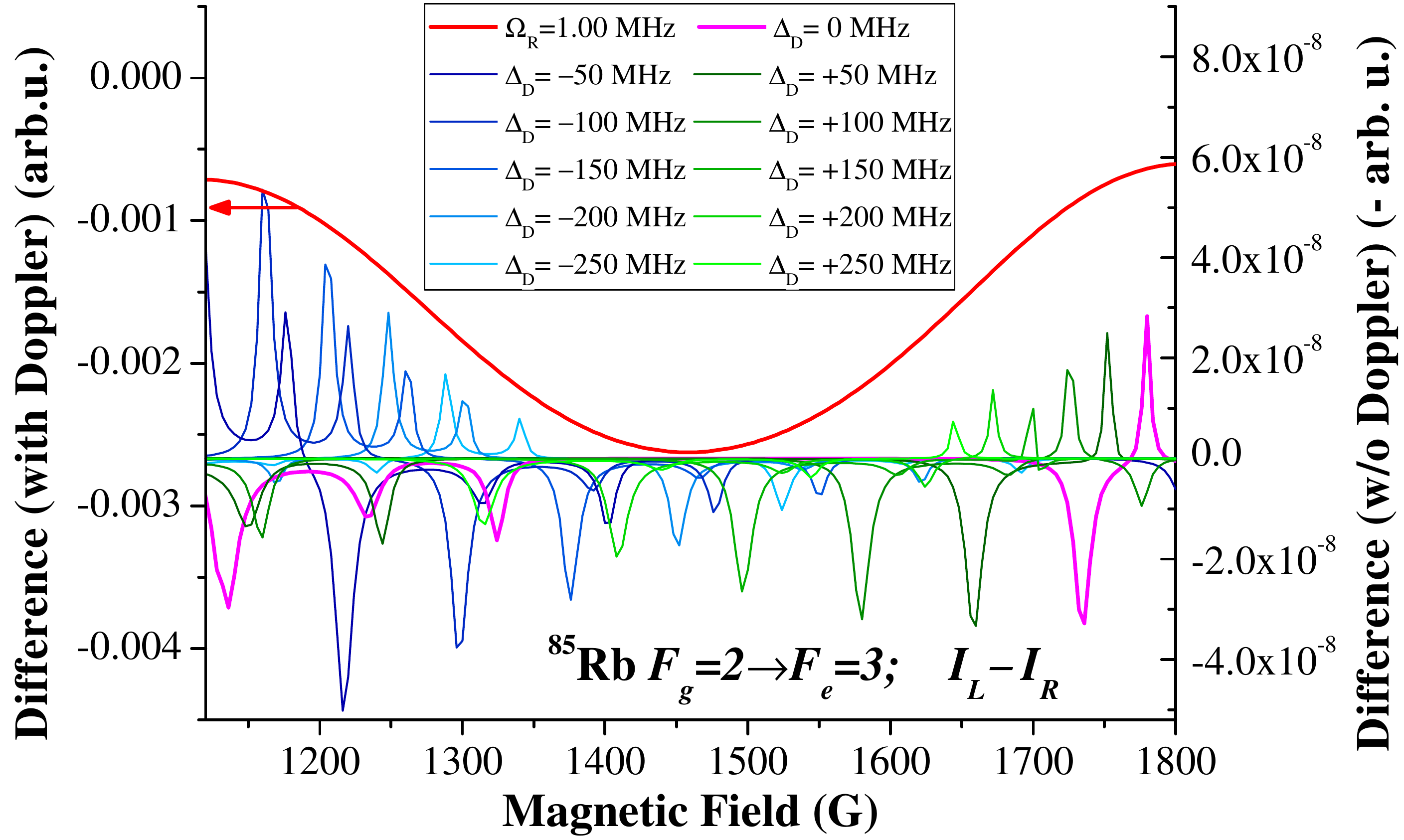}
    \centering
    \caption{\label{fig:im9}Red curve (left axis): theoretical data of the difference between two circularly polarized light components ($I_L-I_R$) with averaging over the Doppler profile. Colored curves (right axis): theoretical data of the difference between two circularly polarized light components without averaging over the Doppler profile.  Different colors represent various velocity groups from the Doppler profile. Magenta: central velocity group; blue curves: negative velocity shift; green curves: positive velocity shift.}
\end{figure}

Figure~\ref{fig:im8} shows the signal dependence on laser power density for the case when the laser frequency was fixed to the $F_g=2\rightarrow F_e=3$ transition of the $^{85}$Rb. As the laser power is being increased, the aforementioned change of the sign of circularity disappears for laser power densities greater than 1.78~mW/cm$^2$ i.e. for all magnetic field values the circularity stays negative. In order to understand the root cause of the circularity lineshapes, theoretical simulations omitting the averaging over the Doppler profile were carried out for various velocity groups of the Doppler profile (Fig.~\ref{fig:im9}). The pronounced peak (structure) seen in Figure~\ref{fig:im8}d at approximately 1500~G would appear to be connected with magnetic sublvels coming into resonance with the laser light, but the LIF component signals in Figure~\ref{fig:im8}a-b-c clearly show a minimum at 1500~G. The origin of this non-zero circularity can be understood by looking at Figure~\ref{fig:im9}: as different velocity groups come into resonance some shift the transverse orientation of the angular momentum in the positive direction and some in the negative. When the summation over all of these contributing velocity groups are combined, only the ones that were not compensated by other velocity groups contribute to the observed circularity signal. The arbitrary units in both vertical axis in Figure~\ref{fig:im9} are directly comparable as the ones on the left correspond to the red line, which is the LIF signal difference obtained by adding the difference signals from separate velocity groups multiplied by the corresponding factor from the Doppler profile.

\begin{figure} 
    \includegraphics[width=1\linewidth]{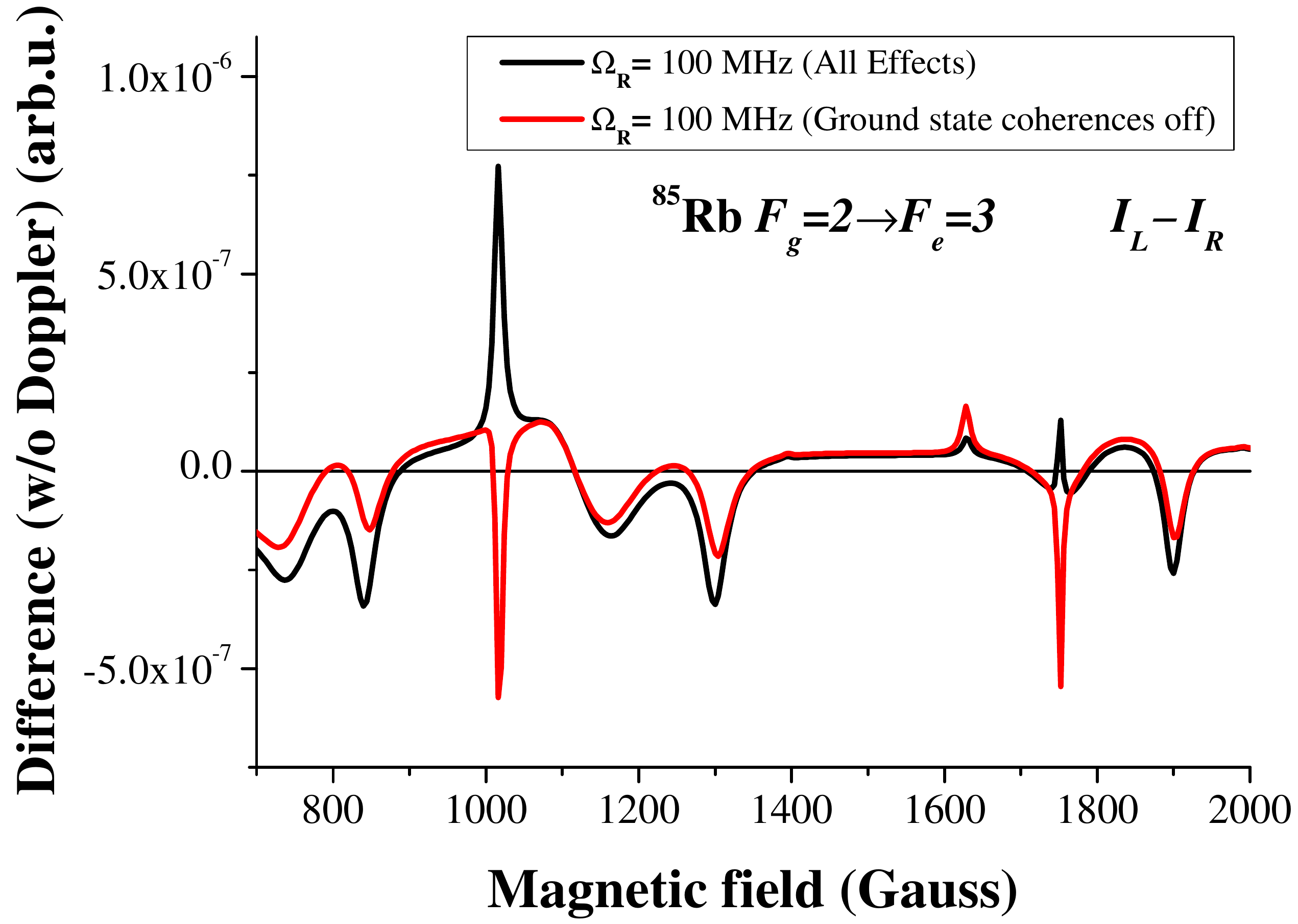}
    \centering
    \caption{\label{fig:im10}$I_L-I_R$ dependence on the magnetic field from central velocity group -- averaging over the Doppler profile is omitted. Red curve: theoretical data from $^{85}$Rb with Rabi frequency 100~MHz with large $\gamma_{non-diagonal}$; Black curve: theoretical data from $^{85}$Rb with Rabi frequency 100~MHz with normal $\gamma_{non-diagonal}$.}
\end{figure}

As the transverse angular momentum AOC is a coherent effect, we wanted to distinguish between the ground-state coherent effects and the excited-state coherent effects contributing to the signal. At low Rabi frequencies the effect of ground-state coherence transfer was minute, and because the underlying causes for the signal shapes can be better understood by analysing the LIF signals coming from separate velocity groups (omitting the averaging over the Doppler profile), we show the simulated curves in Figure~\ref{fig:im10} for the central velocity group (with respect to the exciting laser frequency) with large Rabi frequency. 

With the aim to distinguish which features in the signal are caused by ground-state coherent effects, we set the non-diagonal density matrix elements (in Eq.~\eqref{eq:zcgg}) to zero. We did this by increasing the relaxation rate $\gamma_{non\mhyphen diagonal}$ of only these elements with the ratio of $\gamma_{non\mhyphen diagonal}/\gamma_{diagonal} = 10^9$ with respect to the $\gamma_{diagonal}$ which is the normal transit relaxation rate experienced by diagonal elements. This allowed us to observe the influence of transfer of coherences from the ground state to the excited state. Figure~\ref{fig:im10} shows the comparison of the two cases of simulated LIF signals from the central velocity group. The red curve (in Fig.~\ref{fig:im10}) corresponds to the case when the ground-state coherent effects were set to zero whereas the black curve -- when all the elements in the density matrix experience normal relaxation.

As can be seen from the differences in the two curves (Fig.~\ref{fig:im10}), some features, e.g. at approximately 1300~G, persist in both curves virtually unchanged, but some features experience a dramatic change e.g. features at 1000~G and 1750~G indicating that these features are directly connected to the ground-state coherent effects. Both features show a change in the direction of angular momentum orientation -- when the ground-state coherences were set to zero $I_L<I_R$, while the black curve shows the signal to be $I_L>I_R$ when the parameters for all effects were set to normal values.
When the averaging over the Doppler profile is included, these features become less pronounced as the signals from different velocity groups compensate each other, causing the overall signal to approach zero (much like in the analysis of Figure~\ref{fig:im9}).
This is partially verified by experimentally observed signals -- the feature at 1000~G and 1750~G (see Fig.~\ref{fig:im8}f) also exhibits a tendency of increase of $I_L-I_R$ signal.

\section{Conclusion}

When the coherent effects in the manifold of atomic angular momentum magnetic sublevels, induced by interaction of atoms with laser radiation, are conceptually discussed, very often the primer attention is paid to the creation of coherent superposition of these sublevels due to two factors. First, exctiaton light polarization components capable to excite coherently these sublevels are considered and, second, transition probabilities determined by the transition dipole moments between angular momentum states are accounted for \cite{Auzinsh:2010book}. 

In this paper we analyze in detail and show that on top of these effects a very important role in this process is played by the magnetic scanning of the magnetic sublevels in the external magnetic field. In the Paschen--Back effect regime it leads to two effects: first, nonlinear magnetic sublevel splitting that can lead to angular momentum spatial distribution symmetry breaking -- the alignment-to-orientation conversion and, second, it causes changes in the transition probabilities due to magnetic sublevel mixing the magnetic field.

And a second very important moment in the analysis of laser light--atom interaction is a necessity for clear separation of incoherent (related to the populations distribution of magnetic sublevels) and coherent (determined by a well defined phase relations of magnetic sublevel wave functions) contributions to the observed signals.

In this paper we have shown that, for example, in Rb atoms used in this study, at a different magnetic field strength not only different hyperfine transitions of a specific isotope of an atom are coming into resonance with laser radiation, but the same laser radiation at different magnetic field strength can excite hyperfine transitions in different isotopes of rubidium atoms. This effect appears due to magnetic sublevel scanning and primarily is incoherent effect, see Fig.~\ref{fig:im5} and the analysis of it.

And, finally, based on the comparison of signals obtained in numerical model in which we are able to ``switch-off" and ``switch-on" different relaxation processes,we managed to get evidence that specific features in the observed signals are  determined by the alignment-to-orientation conversion in the atomic ground state, see Figs. \ref{fig:im8} and \ref{fig:im10} and analysis there. We believe that alignment-to-orientation conversion in the ground state of atoms has not been identified before. The clear understanding of the presence of these effects is important for applications as well as for research in fundamental physics in the table top atomic physics experiments, for example, in search of the permanent electric dipole moment of an electron -- EDM experiments.

\begin{acknowledgments}
A. Mozers acknowledges support from ERAF PostDoc Latvia project No. 1.1.1.2/16/117 "Experimental and theoretical signals of ground-state angular momentum alignment-to-orientation conversion by the influence of laser radiation and external magnetic field in atomic alkali metal vapour". The authors are thankful to Dr. L.~Kalvans for discussions and advice with LIF signal simulations.
\end{acknowledgments}

\bibliography{main_AOC_D1}

\end{document}